\documentclass[aps,reprint]{revtex4}

\usepackage{graphicx}
\usepackage{textcomp}
\usepackage{ulem}

\begin{document}

\title{On orbit performance of the GRACE Follow-On Laser Ranging Interferometer} 

\author{Klaus Abich } \affiliation{DLR Institut f\"ur Raumfahrtsysteme, Robert-Hooke-Str. 7, 28359 Bremen, Germany}
\author{Alexander Abramovici } \affiliation{Jet Propulsion Laboratory, California Institute of Technology, 4800 Oak Grove Drive Pasadena \mbox{CA 91109,~USA}}
\author{Bengie Amparan } \affiliation{Ball Aerospace and Technologies Corporation, \mbox{PO Box 1062, Boulder, CO 80306,~USA}}
\author{Andreas Baatzsch } \affiliation{SpaceTech GmbH, Seelbachstrasse 13, 88090 Immenstaad, Germany}
\author{Brian Bachman Okihiro } \affiliation{Jet Propulsion Laboratory, California Institute of Technology, 4800 Oak Grove Drive Pasadena \mbox{CA 91109,~USA}}
\author{David C. Barr } \affiliation{Jet Propulsion Laboratory, California Institute of Technology, 4800 Oak Grove Drive Pasadena \mbox{CA 91109,~USA}}
\author{Maxime P. Bize } \affiliation{Jet Propulsion Laboratory, California Institute of Technology, 4800 Oak Grove Drive Pasadena \mbox{CA 91109,~USA}}
\author{Christina Bogan } \altaffiliation{now at Volkswagen AG, Wolfsburg, Germany} \affiliation{Max-Planck-Institut f\"ur Gravitationsphysik (Albert-Einstein-Institut) and Institut f\"ur Gravitationsphysik of Leibniz Universit\"at Hannover, Callinstrasse 38,   \mbox{30167 Hannover, Germany}} 
\author{Claus Braxmaier } \affiliation{DLR Institut f\"ur Raumfahrtsysteme, Robert-Hooke-Str. 7, 28359 Bremen, Germany}
\author{Michael J. Burke } \affiliation{Jet Propulsion Laboratory, California Institute of Technology, 4800 Oak Grove Drive Pasadena \mbox{CA 91109,~USA}}
\author{Ken C. Clark } \affiliation{Jet Propulsion Laboratory, California Institute of Technology, 4800 Oak Grove Drive Pasadena \mbox{CA 91109,~USA}}
\author{Christian Dahl } \affiliation{SpaceTech GmbH, Seelbachstrasse 13, 88090 Immenstaad, Germany}
\author{Katrin Dahl } \affiliation{SpaceTech GmbH, Seelbachstrasse 13, 88090 Immenstaad, Germany}
\author{Karsten Danzmann } \affiliation{Max-Planck-Institut f\"ur Gravitationsphysik (Albert-Einstein-Institut) and Institut f\"ur Gravitationsphysik of Leibniz Universit\"at Hannover, Callinstrasse 38,   \mbox{30167 Hannover, Germany}} 
\author{Mike A. Davis } \affiliation{Ball Aerospace and Technologies Corporation, \mbox{PO Box 1062, Boulder, CO 80306,~USA}}
\author{Glenn de Vine } \affiliation{Jet Propulsion Laboratory, California Institute of Technology, 4800 Oak Grove Drive Pasadena \mbox{CA 91109,~USA}}
\author{Jeffrey A. Dickson } \affiliation{Jet Propulsion Laboratory, California Institute of Technology, 4800 Oak Grove Drive Pasadena \mbox{CA 91109,~USA}}
\author{Serge Dubovitsky } \affiliation{Jet Propulsion Laboratory, California Institute of Technology, 4800 Oak Grove Drive Pasadena \mbox{CA 91109,~USA}}
\author{Andreas Eckardt } \affiliation{DLR  Institut f\"ur Optische Sensorsysteme, Rutherfordstrasse 2, \mbox{12489 Berlin-Adlershof, Germany}}
\author{Thomas Ester } \affiliation{Tesat-Spacecom GmbH \& Co KG, Gerberstr. 49, 71522 Backnang, Germany}
\author{Germ\'an Fern\'andez Barranco } \affiliation{Max-Planck-Institut f\"ur Gravitationsphysik (Albert-Einstein-Institut) and Institut f\"ur Gravitationsphysik of Leibniz Universit\"at Hannover, Callinstrasse 38,   \mbox{30167 Hannover, Germany}} 
\author{Reinhold Flatscher } \affiliation{Airbus Defence \& Space, 88039 Friedrichshafen, Germany}
\author{Frank Flechtner} \affiliation{Deutsches GeoForschungsZentrum GFZ, Telegrafenberg, 14473 Potsdam, Germany} \affiliation{Technische Universit\"at Berlin, Strasse des 17. Juni 135, 10623 Berlin, Germany}
\author{William M. Folkner } \affiliation{Jet Propulsion Laboratory, California Institute of Technology, 4800 Oak Grove Drive Pasadena \mbox{CA 91109,~USA}}
\author{Samuel Francis } \affiliation{Jet Propulsion Laboratory, California Institute of Technology, 4800 Oak Grove Drive Pasadena \mbox{CA 91109,~USA}}
\author{Martin S. Gilbert } \affiliation{Jet Propulsion Laboratory, California Institute of Technology, 4800 Oak Grove Drive Pasadena \mbox{CA 91109,~USA}}
\author{Frank Gilles } \affiliation{SpaceTech GmbH, Seelbachstrasse 13, 88090 Immenstaad, Germany}
\author{Martin Gohlke } \affiliation{DLR Institut f\"ur Raumfahrtsysteme, Robert-Hooke-Str. 7, 28359 Bremen, Germany}
\author{Nicolas Grossard } \affiliation{iXblue Photonics, 34 rue de la Croix de fer, \mbox{78100 Saint Germain en Laye, France}}
\author{Burghardt Guenther } \affiliation{DLR  Institut f\"ur Optische Sensorsysteme, Rutherfordstrasse 2, \mbox{12489 Berlin-Adlershof, Germany}}
\author{Philipp Hager } \altaffiliation{now at ESA ESTEC,  Noordwijk, Netherlands} \affiliation{SpaceTech GmbH, Seelbachstrasse 13, 88090 Immenstaad, Germany}
\author{Jerome Hauden } \affiliation{iXblue Photonics, 34 rue de la Croix de fer, \mbox{78100 Saint Germain en Laye, France}}
\author{Frank Heine } \affiliation{Tesat-Spacecom GmbH \& Co KG, Gerberstr. 49, 71522 Backnang, Germany}
\author{Gerhard Heinzel }  \email[{Contact address:} ]{gerhard.heinzel@aei.mpg.de} \affiliation{Max-Planck-Institut f\"ur Gravitationsphysik (Albert-Einstein-Institut) and Institut f\"ur Gravitationsphysik of Leibniz Universit\"at Hannover, Callinstrasse 38,   \mbox{30167 Hannover, Germany}} 
\author{Mark Herding } \affiliation{SpaceTech GmbH, Seelbachstrasse 13, 88090 Immenstaad, Germany}
\author{Martin Hinz } \affiliation{Hensoldt Optronics GmbH, Carl-Zeiss-Stra{\ss}e 22, 73447 Oberkochen, Germany}
\author{James Howell } \affiliation{Ball Aerospace and Technologies Corporation, \mbox{PO Box 1062, Boulder, CO 80306,~USA}}
\author{Mark Katsumura } \affiliation{Jet Propulsion Laboratory, California Institute of Technology, 4800 Oak Grove Drive Pasadena \mbox{CA 91109,~USA}}
\author{Marina Kaufer } \affiliation{SpaceTech GmbH, Seelbachstrasse 13, 88090 Immenstaad, Germany}
\author{William Klipstein } \affiliation{Jet Propulsion Laboratory, California Institute of Technology, 4800 Oak Grove Drive Pasadena \mbox{CA 91109,~USA}}
\author{Alexander Koch } \affiliation{Max-Planck-Institut f\"ur Gravitationsphysik (Albert-Einstein-Institut) and Institut f\"ur Gravitationsphysik of Leibniz Universit\"at Hannover, Callinstrasse 38,   \mbox{30167 Hannover, Germany}}
\author{Micah Kruger } \affiliation{Ball Aerospace and Technologies Corporation, \mbox{PO Box 1062, Boulder, CO 80306,~USA}}
\author{Kameron Larsen } \affiliation{Jet Propulsion Laboratory, California Institute of Technology, 4800 Oak Grove Drive Pasadena \mbox{CA 91109,~USA}}
\author{Anton Lebeda } \affiliation{APCON AeroSpace \& Defence, Prof.~Messerschmitt-Str.~10, \mbox{85579 Neubiberg, Germany} } 
\author{Arnold Lebeda } \affiliation{APCON AeroSpace \& Defence, Prof.~Messerschmitt-Str.~10, \mbox{85579 Neubiberg, Germany} }
\author{Thomas Leikert } \affiliation{Hensoldt Optronics GmbH, Carl-Zeiss-Stra{\ss}e 22, 73447 Oberkochen, Germany}
\author{Carl Christian Liebe } \affiliation{Jet Propulsion Laboratory, California Institute of Technology, 4800 Oak Grove Drive Pasadena \mbox{CA 91109,~USA}}
\author{Jehhal Liu } \affiliation{Jet Propulsion Laboratory, California Institute of Technology, 4800 Oak Grove Drive Pasadena \mbox{CA 91109,~USA}}
\author{Lynette Lobmeyer} \affiliation{Ball Aerospace and Technologies Corporation, \mbox{PO Box 1062, Boulder, CO 80306,~USA}}
\author{Christoph Mahrdt } \altaffiliation{now at Siemens Mobility GmbH, Braunschweig, Germany}\affiliation{Max-Planck-Institut f\"ur Gravitationsphysik (Albert-Einstein-Institut) and Institut f\"ur Gravitationsphysik of Leibniz Universit\"at Hannover, Callinstrasse 38,   \mbox{30167 Hannover, Germany}} 
\author{Thomas Mangoldt } \affiliation{DLR  Institut f\"ur Optische Sensorsysteme, Rutherfordstrasse 2, \mbox{12489 Berlin-Adlershof, Germany}}
\author{Kirk McKenzie } \email[{Contact address:} ]{kirk.mckenzie@jpl.nasa.gov} \affiliation{Jet Propulsion Laboratory, California Institute of Technology, 4800 Oak Grove Drive Pasadena \mbox{CA 91109,~USA}}
\author{Malte Misfeldt } \affiliation{Max-Planck-Institut f\"ur Gravitationsphysik (Albert-Einstein-Institut) and Institut f\"ur Gravitationsphysik of Leibniz Universit\"at Hannover, Callinstrasse 38,   \mbox{30167 Hannover, Germany}} 
\author{Phillip R. Morton } \affiliation{Jet Propulsion Laboratory, California Institute of Technology, 4800 Oak Grove Drive Pasadena \mbox{CA 91109,~USA}}
\author{Vitali M\"uller } \affiliation{Max-Planck-Institut f\"ur Gravitationsphysik (Albert-Einstein-Institut) and Institut f\"ur Gravitationsphysik of Leibniz Universit\"at Hannover, Callinstrasse 38,   \mbox{30167 Hannover, Germany}} 
\author{Alexander T. Murray } \affiliation{Jet Propulsion Laboratory, California Institute of Technology, 4800 Oak Grove Drive Pasadena \mbox{CA 91109,~USA}}
\author{Don J. Nguyen } \affiliation{Jet Propulsion Laboratory, California Institute of Technology, 4800 Oak Grove Drive Pasadena \mbox{CA 91109,~USA}}
\author{Kolja Nicklaus } \affiliation{SpaceTech GmbH, Seelbachstrasse 13, 88090 Immenstaad, Germany}
\author{Robert Pierce } \affiliation{Ball Aerospace and Technologies Corporation, \mbox{PO Box 1062, Boulder, CO 80306,~USA}}
\author{Joshua A. Ravich } \affiliation{Jet Propulsion Laboratory, California Institute of Technology, 4800 Oak Grove Drive Pasadena \mbox{CA 91109,~USA}}
\author{Gretchen Reavis } \affiliation{Ball Aerospace and Technologies Corporation, \mbox{PO Box 1062, Boulder, CO 80306,~USA}}
\author{Jens Reiche } \affiliation{Max-Planck-Institut f\"ur Gravitationsphysik (Albert-Einstein-Institut) and Institut f\"ur Gravitationsphysik of Leibniz Universit\"at Hannover, Callinstrasse 38,   \mbox{30167 Hannover, Germany}} 
\author{Josep Sanjuan } \affiliation{DLR Institut f\"ur Raumfahrtsysteme, Robert-Hooke-Str. 7, 28359 Bremen, Germany}
\author{Daniel Sch\"utze } \altaffiliation{now at OHB-Systems AG, Bremen, Germany}\affiliation{Max-Planck-Institut f\"ur Gravitationsphysik (Albert-Einstein-Institut) and Institut f\"ur Gravitationsphysik of Leibniz Universit\"at Hannover, Callinstrasse 38,   \mbox{30167 Hannover, Germany}} 
\author{Christoph Seiter } \affiliation{Tesat-Spacecom GmbH \& Co KG, Gerberstr. 49, 71522 Backnang, Germany}
\author{Daniel Shaddock } \altaffiliation{now at Liquid Instruments, Canberra Australia}\affiliation{Jet Propulsion Laboratory, California Institute of Technology, 4800 Oak Grove Drive Pasadena \mbox{CA 91109,~USA}} 
\author{Benjamin Sheard } \altaffiliation{now at OHB-Systems AG, Bremen, Germany}\affiliation{Max-Planck-Institut f\"ur Gravitationsphysik (Albert-Einstein-Institut) and Institut f\"ur Gravitationsphysik of Leibniz Universit\"at Hannover, Callinstrasse 38,   \mbox{30167 Hannover, Germany}} 
\author{Michael Sileo } \affiliation{Ball Aerospace and Technologies Corporation, \mbox{PO Box 1062, Boulder, CO 80306,~USA}}
\author{Robert Spero } \affiliation{Jet Propulsion Laboratory, California Institute of Technology, 4800 Oak Grove Drive Pasadena \mbox{CA 91109,~USA}}
\author{Gary Spiers } \affiliation{Jet Propulsion Laboratory, California Institute of Technology, 4800 Oak Grove Drive Pasadena \mbox{CA 91109,~USA}}
\author{Gunnar Stede } \altaffiliation{now at Leibnizschule, Hannover, Germany}\affiliation{Max-Planck-Institut f\"ur Gravitationsphysik (Albert-Einstein-Institut) and Institut f\"ur Gravitationsphysik of Leibniz Universit\"at Hannover, Callinstrasse 38,   \mbox{30167 Hannover, Germany}} 
\author{Michelle Stephens } \altaffiliation{now at NIST, Boulder, USA} \affiliation{Ball Aerospace and Technologies Corporation, \mbox{PO Box 1062, Boulder, CO 80306,~USA}}
\author{Andrew Sutton } \affiliation{Jet Propulsion Laboratory, California Institute of Technology, 4800 Oak Grove Drive Pasadena \mbox{CA 91109,~USA}}
\author{Joseph Trinh } \affiliation{Jet Propulsion Laboratory, California Institute of Technology, 4800 Oak Grove Drive Pasadena \mbox{CA 91109,~USA}}
\author{Kai Voss } \affiliation{SpaceTech GmbH, Seelbachstrasse 13, 88090 Immenstaad, Germany}
\author{Duo Wang } \affiliation{Jet Propulsion Laboratory, California Institute of Technology, 4800 Oak Grove Drive Pasadena \mbox{CA 91109,~USA}}
\author{Rabi T. Wang } \affiliation{Jet Propulsion Laboratory, California Institute of Technology, 4800 Oak Grove Drive Pasadena \mbox{CA 91109,~USA}}
\author{Brent Ware } \affiliation{Jet Propulsion Laboratory, California Institute of Technology, 4800 Oak Grove Drive Pasadena \mbox{CA 91109,~USA}}
\author{Henry Wegener } \affiliation{Max-Planck-Institut f\"ur Gravitationsphysik (Albert-Einstein-Institut) and Institut f\"ur Gravitationsphysik of Leibniz Universit\"at Hannover, Callinstrasse 38,   \mbox{30167 Hannover, Germany}} 
\author{Steve Windisch } \affiliation{Tesat-Spacecom GmbH \& Co KG, Gerberstr. 49, 71522 Backnang, Germany}
\author{Christopher Woodruff } \affiliation{Jet Propulsion Laboratory, California Institute of Technology, 4800 Oak Grove Drive Pasadena \mbox{CA 91109,~USA}}
\author{Bernd Zender } \affiliation{DLR  Institut f\"ur Optische Sensorsysteme, Rutherfordstrasse 2, \mbox{12489 Berlin-Adlershof, Germany}}
\author{Marcus Zimmermann } \affiliation{Hensoldt Optronics GmbH, Carl-Zeiss-Stra{\ss}e 22, 73447 Oberkochen, Germany}

\date{\today}

\begin{abstract} 
  The Laser Ranging Interferometer (LRI) instrument on the Gravity Recovery and Climate Experiment (GRACE) Follow-On mission has provided the first laser interferometeric range measurements between remote 
spacecraft, separated by approximately 220\,km. 
  Autonomous controls that lock the laser frequency to a cavity reference and establish the 5 degree of freedom two-way laser link between remote spacecraft succeeded on the first attempt. Active beam pointing based on differential wavefront sensing compensates spacecraft attitude fluctuations. The LRI has operated continuously without breaks in phase tracking for more than 50 days, and has shown biased range measurements similar to the primary ranging instrument based on microwaves, but with much less noise at a level of $1\,{\rm nm}/\sqrt{\rm Hz}$ at Fourier frequencies above 100\,mHz.
\end{abstract}

\maketitle

\section{Introduction}
\label{sec:intro}
The Gravity Recovery and Climate Experiment (GRACE) mission \cite{RN1328,RN1326}, in orbit from 2002--2017, was a US-German collaboration that has revolutionized the measurement of the time-variable Earth gravity field. The primary instrument was the Microwave Instrument (MWI) that tracked the variations of the distance (biased range) between the two satellites as they followed each other with 200\,km separation on a near polar orbit about 450\,km above the Earth. These variations reflect the structure of the gravity field and additionally contain small non-gravitational accelerations.

Data processing of the range variations that also involved the on-board accelerometer, GPS-based precise orbit determination, modeling of ocean and solid Earth tides and other, smaller effects yielded as the main result monthly maps of the Earth’s gravity field. GRACE data has been used by hundreds of researchers worldwide in the publication of thousands of papers in climate research, tracking changes in ice and ground water, in geophysics and in many other fields of research.  Desire to continue the 15~years of observations led to GRACE Follow-On -- which is again a US-German partnership and in key aspects a re-build of GRACE with again the MWI as main instrument\cite{RN1315,RN1333}. The main new feature of GRACE Follow-On is the addition of the Laser Ranging Interferometer (LRI) instrument as a technology demonstrator for future Earth-science missions. The LRI also serves as a useful demonstrator for  the Laser Interferometer Space Antenna (LISA)\cite{LISA-ESA}. It measures the same variations in the inter-spacecraft distance as the MWI, but with less noise\cite{RN1317,RN1305}.

These variations have amplitudes of a few 100\,m at the orbital frequency. They contain non-gravitational disturbances and the gravity signal that spans a huge dynamic range between $10^{-8}\,{\rm m}/\sqrt{\rm Hz}$ and $1\,{\rm m}/\sqrt{\rm Hz}$ (Figures 4 and 6 below). Most of that signal encodes the static Earth's gravity field. The temporal variations are much smaller\cite{essd-3-19-2011}, motivating the need for ranging noise in the nm/$\sqrt{\rm Hz}$ range. 

The LRI is the first laser interferometer to be operated between distant satellites. GRACE Follow-On was launched on May 22, 2018 and the first attempt to turn on the LRI took place on June 14, with immediate success.
This article summarizes the design of the LRI in Section \ref{sec:components}, Section \ref{sec:acquisition} discusses the acquisition strategy, and Section~\ref{sec:results} reports the first ranging measurements.

\section{LRI components and design}
\label{sec:components}

\begin{figure*}
\includegraphics[width=0.9\textwidth]{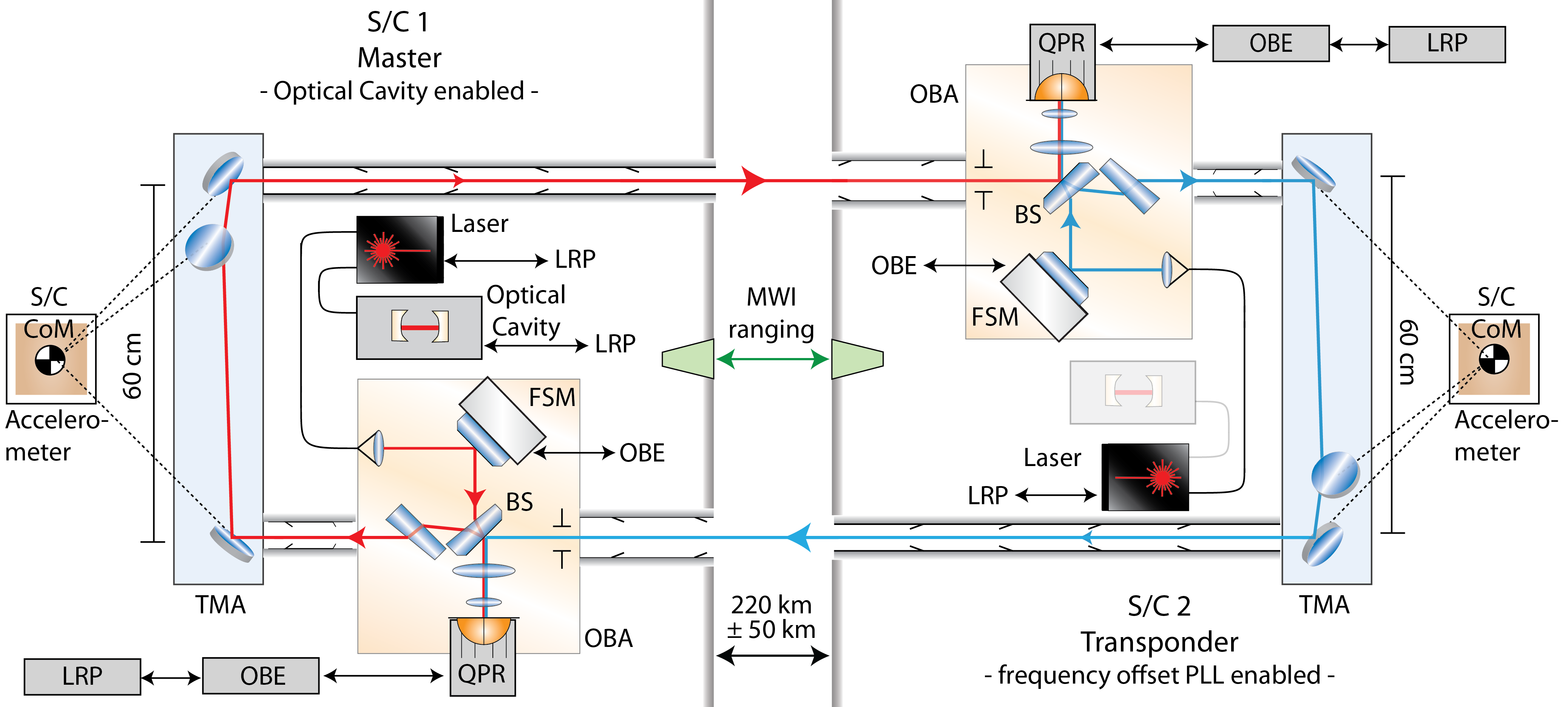}
\caption{\label{fig:sketch}Functional overview of the LRI units on both spacecraft. The LRI units include the laser, cavity, laser ranging processor (LRP), optical bench electronics (OBE), triple mirror assembly (TMA) and optical bench assembly (OBA) with a fast steering mirror (FSM).}
\end{figure*}

Figure 1 shows a functional diagram of the LRI on the two identical spacecraft. Its components include the laser, the cavity, the laser ranging processor (LRP), the optical bench electronics (OBE), the optical bench assembly (OBA), and the triple mirror assembly (TMA). The basic design of the LRI has been described in a separate paper\cite{RN1317}.

The LRI is a US-German cooperative project led by NASA/JPL in the US and the AEI in Hannover, Germany. The laser, cavity, and LRP were contributed by the US, while the TMA, OBA, and OBE were contributed by Germany.

The lasers are Nd:YAG nonplanar ring oscillators manufactured by Tesat-Spacecom. They operate at a wavelength of 1064.5\,nm and produce 25\,mW of fiber coupled laser light. The frequency stability of the laser limits the sensitivity of the LRI instrument at high frequencies. Therefore, both spacecraft carry identical optical cavities, one of which is used to stabilize the laser frequency on the spacecraft designated as master. The flight units of the optical cavities were manufactured by Ball Aerospace based on prototypes developed by Ball Aerospace and JPL under a NASA Earth Science grant\cite{RN1319,RN1321}. Each cavity unit includes an optical phase modulator delivered by iXblue. Ground measurements of the respective LRI components showed frequency noise below $30\,{\rm Hz}/\sqrt{\rm Hz}$ at Fourier frequencies of 1\,mHz and above\cite{RN1304}.

The LRP was developed and built by JPL, based on LISA and Earth Science studies\cite{RN1319,RN1321,RN1304,DAS2016}. It processes the photoreceiver signals both by tracking with digital phase-locked loops for the ranging measurements (phasemeter function) and by continuously running a Fast Fourier Transform (FFT) which is used in particular during acquisition. The LRP commands the fast steering mirror in link acquisition and performs closed loop control for differential wavefront sensing when the inter-spacecraft link is active. It controls the laser frequency as either master or transponder.  The LRI's command and telemetry is handled by the LRP.

The LRI optical system consists of the OBA\cite{nicklaus2017optical} built by SpaceTech GmbH (STI) with associated OBE built by the Deutsches Zentrum f\"ur Luft- und Raumfahrt (DLR) in Adlershof, the TMA from STI and Hensoldt/Zeiss, and associated baffles and mechanical parts (from STI). 

On the Optical Bench, which is relatively simple compared to LISA Pathfinder\cite{LTP,LTPOB} or LISA, light from the laser is delivered through an optical single-mode fiber and custom-made collimator to produce a Gaussian beam of 2.5\,mm radius. It is routed via a fast steering mirror (FSM)\cite{RN1380} from Airbus Defence and Space which can steer the beam direction by several mrad in two axes at rates $>100$\,Hz. A beamsplitter (BS) transmits 10\,\% of the light to be used as local oscillator (LO) to enable heterodyne detection of the nW-level received (RX) light. The latter passes through an aperture of 8\,mm diameter before arriving at the other port of the BS. Aperture and steering mirror have the same distance from the BS. A two-lens imaging system images both with a magnification $m=1/8$ on to the InGaAs Quadrant Photoreceiver (QPR) such that a tilt of either FSM or incoming wavefront translates into an $8\times$ magnified tilt on the QPR with virtually no beam walk. The remaining 90\,\% of the local laser light is reflected by the BS and is routed through the TMA, which consists of three plane mirrors, whose planes are precisely orthogonal to each other and intersect in a virtual vertex located at the Center of Mass (COM) of the accelerometer test mass, the reference point for the ranging measurement\cite{RN1311, RN1313,RN1314,RN1316}. The TMA reverses the direction of the beam and directs it towards the other spacecraft and also displaces it by 60\,cm to avoid the vicinity of the central line-of-sight, which is occupied by other equipment. The divergence of the transmitted (TX) light of 140\,\textmu rad causes the beam to expand to about 30\,m radius, from which a tiny fraction (order of nW) is cut out with the RX aperture at the other end.
One spacecraft acts as master, sending out a laser beam with stabilized frequency.  
The other spacecraft acts as transponder and sends back a laser beam
that is phase-locked with a fixed 10\,MHz offset to the weak incoming light.
Both spacecraft are identical and can be commanded to either master or transponder role. The total mass of the LRI units per spacecraft is 25\,kg, and the nominal power consumption is 35\,W.

The incoming light interferes with the LO, resulting in a beat note around 10\,MHz, the phase of which is recorded with the phasemeter function of the LRP.
The orbital motion contains relative velocities along the line of
sight with an amplitude of $\pm0.5\,$m/s, 
giving rise to Doppler shifts of $\pm0.5\,$MHz, which also contain the gravity signal of interest \cite{Seeber2003}.
On the transponder spacecraft, the offset-phase lock impresses that
Doppler shift on the re-transmitted light, which picks up one more
Doppler shift on its way back. As a result, the phasemeter on the
transponder records a constant 10\,MHz tone, while the beat note on
the master additionally contains twice the Doppler shift.

\begin{figure}
\includegraphics[width=0.4\textwidth]{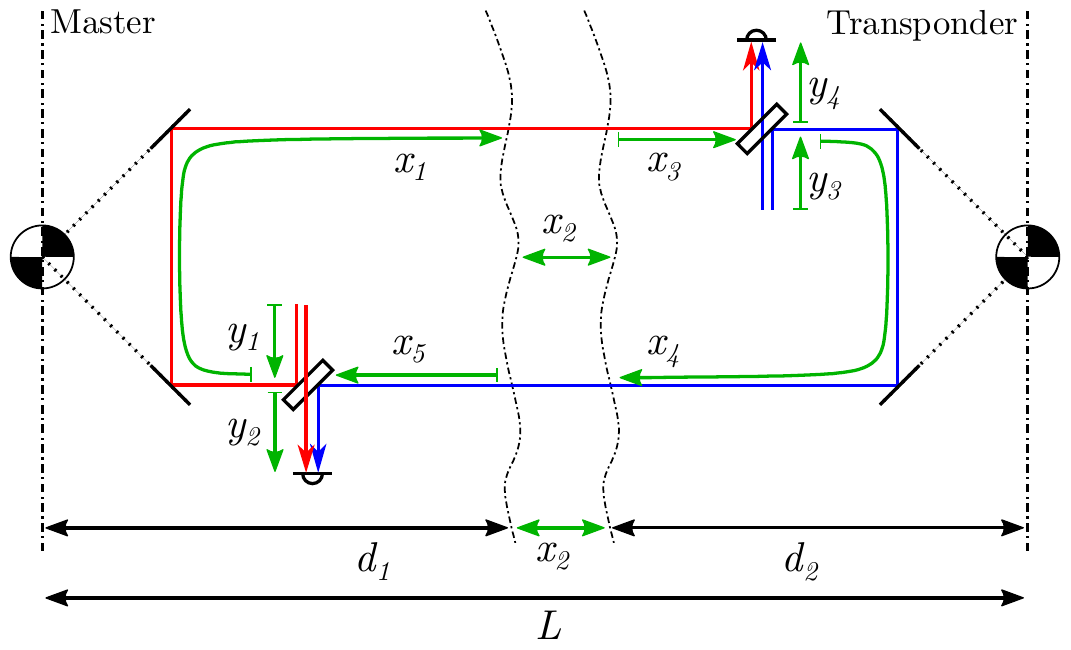}
\caption{\label{fig:lengths3}LRI measured lengths.}
\end{figure}

This is different from the dual one-way ranging system employed by the MWI, but similar to the laser interferometry envisaged for LISA\cite{thorpe2010lisa}. 
An analysis of the optical pathlengths shown in Figure~\ref{fig:lengths3} reveals that the effectively measured quantity is $x_1+x_2+x_3+x_4+x_2+x_5=2(d_1+d_2+x_2)=2L$, i.e.\ twice the separation $L$ of the two TMA vertices, while all other lengths ($y_1$ \dots $y_4$ as well as the delivery fibers etc.) cancel to first order\cite{RN1317}.

A specific feature of the LRI is its beam pointing function\cite{RN1310}, which is required because of the narrower beam width of 140\,\textmu rad half-cone angle compared to the microwave beam of 30\,mrad.

The LRI transmit beam  must point to the other
distant spacecraft with better than 100\,\textmu rad accuracy to
ensure that enough light arrives at the distant RX aperture. Similarly, the weak RX beam must be aligned to the LO beam to about the same accuracy in order to achieve sufficient heterodyne efficiency. The attitude fluctuations of the spacecraft with respect to the line of sight are up to three times larger than that, thus requiring active compensation. Accurate beam pointing is achieved by a single actuator, the FSM, together with Differential Wavefront Sensing \cite{DWS1} signals and the retroreflector properties of the TMA\cite{RN1310,RN1311}.

The requirement for the LRI ranging noise was set to 80\,nm$/\sqrt{\rm
  Hz}$ between 2\,mHz and 100\,mHz, with a noise-shape function of (Figure~\ref{fig:spectrum}) 
\[
\tilde{s}<80\,{\rm nm}/\sqrt{\rm Hz}\sqrt{1+\left(\frac{3\,{\rm mHz}}{f}\right)^2}\sqrt{1+\left(\frac{10\,{\rm mHz}}{f}\right)^2}.
\]
This is one order of magnitude better than the instrumental noise of the MWI, and was designed to be low enough such that it is not the limiting noise source at any frequency. The gravity field resolution is expected to be limited by other noise sources such as tide modeling and aliasing\cite{RN1305}. 

The LRI sensitivity was designed to be limited at higher frequencies by laser frequency noise scaled with the $\approx 220$\,km separation, and at lower frequencies by tilt-to-length coupling originating from the spacecraft attitude jitter in combination with coupling factors of a few 100\,\textmu m/rad originating from alignment tolerances.

The LRI instrument operates autonomously and produces continuous range telemetry  at a $\approx 10$\,samples per second update rate to be used for gravity field recovery. 

The similarities to the planned LISA mission include the measurement band (mHz), the heterodyne frequency (MHz), the received light power level (nW), the offset locked transponder scheme, the digital phase-locked loop phasemeter, the laser frequency stability achieved by a stable external cavity, autonomous locking procedures for the cavity and the transponder loop, the quadrant photoreceivers with differential wavefront sensing, and the principle of the five degree of freedom acquisition procedure. In spite of remaining differences, this first successful demonstration of long distance spacecraft interferometry is thus highly relevant for the ongoing design of LISA as well. 

\section{Link acquisition}
\label{sec:acquisition}

Early  in the development of the LRI, link acquisition was identified as a critical function and it has therefore been intensively studied\cite{Mahrdt2014,RN1312,RN1307,Koch2018}. Link acquisition requires that five degrees of freedom are simultaneously  within narrow ranges around their optimum. These are two pointing angles (pitch and yaw) per spacecraft, each of which has to be within $\pm100\,$\textmu rad, and the difference between the two absolute laser frequencies which has to fulfill $|\Delta f|<15$\,MHz due to the photoreceiver and phasemeter bandwidths. The unknown offsets in these quantities are approximately 10 times larger after launch. In the initial acquisition scan, an exhaustive scan of the 5-D parameter space is performed over the course of several hours. Its purpose is to determine the static pointing offsets between the LRI optical axes and the nominal spacecraft pointing direction, and the frequency difference between the two lasers.  
The LRPs on both spacecraft continuously compute FFT spectra of the QPR samples ($N=4096$ at $f_{\rm samp} \approx 39\,{\rm MHz}$), resulting in power measurements roughly every 100\,\textmu s for each frequency bin of 10\,kHz width.  Peaks in the spectra are identified and stored. These are downloaded along with the corresponding commanded positions of the FSM and the laser frequency setpoint and analyzed on ground (Figure~\ref{fig:scan}). 

The scan patterns are a slow hexagonal pattern on the master spacecraft and a fast Lissajous pattern on the transponder spacecraft. They are added to the reference direction which is provided by the spacecraft as an estimate of its attitude error. The slowest outermost actuation scans the transponder laser frequency, while the master stays locked on a cavity resonance. The angular scan range is approx.~$\pm$ 3\,mrad. The fast Lissajous pattern is scanned at 100\,Hz and 2\,Hz in the two axes, respectively, while the hexagonal pattern consists of 1200 discrete points with a hold time of 0.56\,s per point, resulting in a duration of 11.7\,minutes for a single hexagonal master scan during the initial acquisition phase. The transponder laser frequency sweeps the 320\,MHz peak-peak uncertainty range with 10.5\,kHz/s, which yields a duration of approx.~8.5\,hr for the full initial acquisition scan.

The downlinked data are processed to determine the angular and frequency offsets. These offsets are uploaded and then the instrument is commanded into re-acquisition mode, which performs a similar 5-D scan, but with a ten times reduced range ($\pm$ 300\,\textmu rad). When the LRI detects a heterodyne signal it stops the scanning and transitions to phase tracking mode. Re-acquisition takes less than 5~minutes.

\section{On-orbit performance}
\label{sec:results}
\begin{figure}
\includegraphics[width=0.5\textwidth, trim=70 0 30 00]{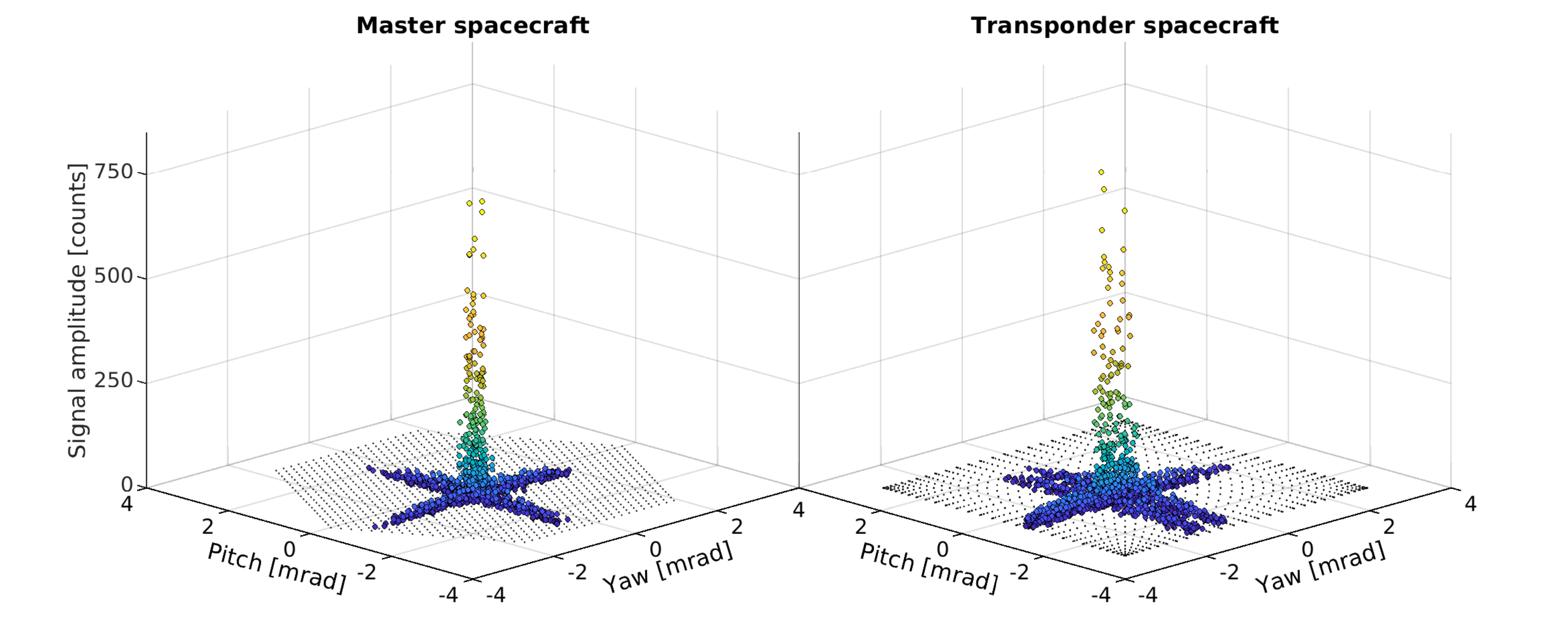}
\caption{\label{fig:scan}FFT amplitude peaks from the initial acquisition scan on June 13, 2018.
 Non-zero values represent the instances ($<$\,1\,sec) during the 8.5\,hour scan when all five degrees of freedom were near their correct values and therefore a heterodyne signal was detected.}
\end{figure}

\begin{figure}
\includegraphics[width=0.5\textwidth, trim=80 80 10 130]{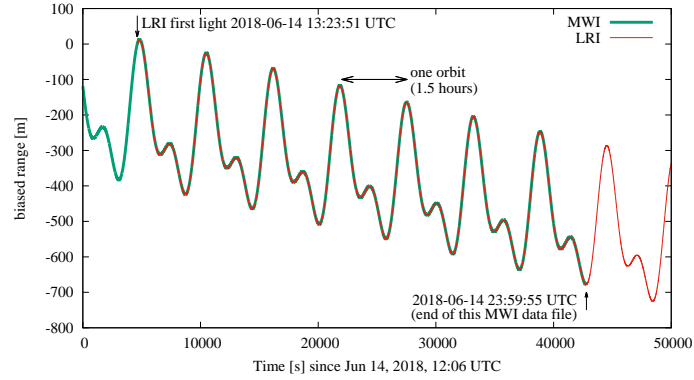}
\caption{\label{fig:lrikbr}First LRI ranging measurements show good agreement with the MWI.}
\end{figure}

GRACE Follow-On was launched on May 22, 2018 by a SpaceX Falcon-9 launch vehicle from the Vandenberg Air Force Base, California, together with five Iridium-Next satellites. Contact was established immediately after separation from the second stage, 11 minutes after launch.  Orbit insertion was nearly perfect. After routine checkout procedures of the spacecraft, the MWI system, and the individual LRI components, controlled from the German Space Operations Center (GSOC) in Oberpfaffenhofen, the initial acquisition scan was started on June 13 2018.  Many peaks of high Signal-to-Noise Ratio were found (Figure~\ref{fig:scan}). The slow scan on the master yielded a single maximum, while the fast Lissajous scan on the transponder produced several maxima due to the delay between FSM-commanded position and FFT peak finding, which was expected and corrected in the immediately performed data analysis on ground. Equally expected were the `+' shaped sidelobes, which stem from the slits in the QPRs\cite{Mahrdt2014}. The resulting pointing offsets as determined by the maxima in  Figure~\ref{fig:scan} were on the order of 0.5~--~1\,mrad.
The calculated position and frequency offsets were uplinked, along with the command to enter re-acquisition mode, on June 14. On the next downlink, both spacecraft reported that they had entered science mode on the first attempt. The LRI stayed in science data collection mode with no interruptions of the links other than intentional mode switches or spacecraft activities unrelated to the LRI. The few instances of re-acquisition  were all immediately successful within a few minutes. The longest segment without any interruption of the link during the first months of operation was 55.5 days long, about 850 orbits. The immediate comparison of LRI range and MWI range is shown in Figure~\ref{fig:lrikbr} and confirms that the LRI measures the biased inter-spacecraft range. While the LRI was designed and verified on ground to operate with a carrier-to-noise ratio (CNR) of as low as 70\,dB-Hz, the observed CNR during initial operation was 88\,dB-Hz, providing ample margin.

\begin{figure}
\includegraphics[width=0.5\textwidth, trim=00 00 10 10]{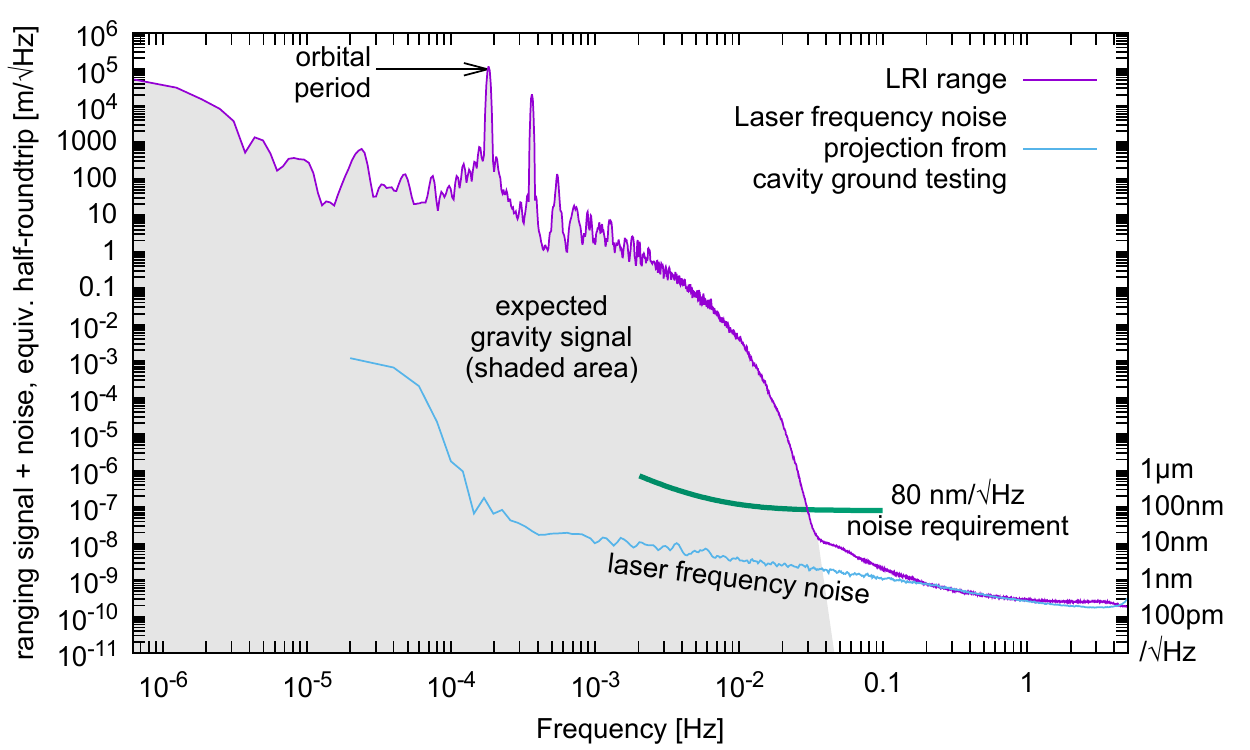}
\caption{\label{fig:spectrum}Amplitude spectral density of LRI ranging measurements.  The purple line shows the ranging signal, after subtraction of phase jumps, which is dominated by the gravity signal below 30\,mHz. The blue line shows the stabilized
 laser frequency noise projected from ground measurements and the green line shows the LRI requirement.}
\end{figure}

\begin{figure}
\includegraphics[width=0.5\textwidth, trim=80 80 10 10]{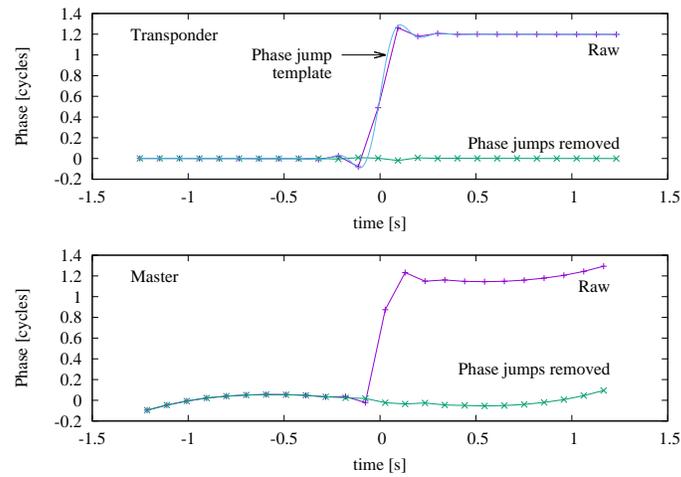}
\caption{\label{fig:glitch}The LRI instrument exhibits occasional phase jumps in the inter-spacecraft range measurement. These are correlated with thruster activations. The analysis shows that phase tracking is always maintained and the phase jumps can always be removed without any impact on the range data. A polynomial of degree 1 and 2 was subtracted from transponder and master phase, respectively, to make the jumps visible in the plot.}
\end{figure}

Figure~\ref{fig:spectrum} shows the amplitude spectral density of the LRI range measurement, where the only processing was the removal of phase jumps (see below). 

It contains the full gravity signal and all non-gravitational observations. At high frequencies, it is limited by laser frequency noise, in good agreement with ground measurements of the cavity performance, which translates to e.g.~$\approx$\ 4\,nm$/\sqrt{\rm Hz}$ at 0.1\,Hz. Towards higher frequencies, the noise level falls off to values as low as 0.2\,nm$/\sqrt{\rm Hz}$. 

The measured phase has occasional jumps coincident with attitude control thruster activation, which occur typically a few times per orbit revolution. These are modeled by a step convolved with the final antialiasing filter response, and removed from the data, with residuals below the noise floor (Figure~\ref{fig:glitch}).  Thus, they have no effect upon the performance of the instrument.

Further analysis of the LRI ranging performance below 0.1\,Hz will be possible from the residuals after the gravity field recovery, which is ongoing work.

\section{Conclusion}

Since initial turn on, the LRI has continuously returned range data, interrupted only by spacecraft operation. Its noise at frequencies that can be evaluated directly is well below the requirement, reaching $10\,{\rm nm}/\sqrt{\rm Hz}$ at 40\,mHz and $300\,{\rm pm}/\sqrt{\rm Hz}$ at 1\,Hz. The in-band noise level will only be accessible after full gravity field analysis. In addition to providing a low-noise laser ranging measurement demonstration for future geodesy missions, the LRI also demonstrates inter-spacecraft interferometry for LISA.

\begin{acknowledgments}
The LRI team would like to acknowledge the GRACE Follow-On Project teams at JPL including Project management, Mission Assurance and Quality Assurance, Project and Flight System engineering, Mission Operations and Science Data Systems at JPL and CSR. It is a pleasure to acknowledge the spacecraft provider: Airbus Defence and Space for their excellence in integrating the LRI instrument to precise requirements without which the LRI could not meet its goals. The LRI team would like to acknowledge the Deutsches Zentrum f\"ur Luft und Raumfahrt (DLR) and German Research Centre for Geosciences (GFZ) Mission Operation teams at the German Space Operation Center (GSOC) for their excellent work and support in commanding the LRI instrument in all mission phases.

The Australian National University developed an alternative Triple mirror assembly and were involved in testing and concept development for link acquisition. 

Part of the research was carried out at the Jet Propulsion Laboratory, California Institute of Technology, under a contract with the National Aeronautics and Space Administration. GRACE-FO is a partnership between NASA and German Research Centre for Geosciences in Potsdam, Germany. JPL manages the mission for NASA’s Science Mission Directorate.

The development of the LRI in Germany was supported by Bundesministerium f\"ur Bildung und Forschung (BMBF), project number 03F0654B, Deutsche
Forschungsgemeinschaft (DFG) and Deutsches Zentrum f\"ur Luft- und Raumfahrt (DLR).

\end{acknowledgments}

\bibliography{LRI}

\end{document}